\documentclass[11pt,twoside]{article}

\usepackage{asp2004}
\usepackage{graphicx}
\usepackage{epsf}
\usepackage{psfig}
\usepackage{lscape}

\begin{document}
\title{A Search for Molecular Gas \\ in Low Luminosity Radio Galaxies}
%%% Fill in title
\author{I. Prandoni$^1$, R.A. Laing$^2$, P. Parma$^1$, H.R. de Ruiter$^3$, 
F.M. Montenegro-Montes$^1$ and T.L. Wilson$^2$}   
%%% Fill in author names
\affil{$^1$INAF - Istituto di Radioastronomia, Via P. Gobetti 101, I-40129 
Bologna, Italy \\
$^2$European Southern Observatory, Karl-Schwarzschild-Str. 2, D-85748 Garching
b. Munch\"en, Germany \\
$^3$INAF - Osservatorio Astronomico di Bologna, Via Ranzani 1, I-40127 
Bologna, Italy}    %%% Fill in author affiliations

\begin{abstract} %%% Abstract to run on from here.
We discuss CO spectral line data of a volume-limited sample of 23 nearby 
($z<0.03$) low luminosity radio galaxies, selected from the B2 catalogue.
Most of such objects (16/23) have HST imaging.\\
Our aim is to establish the
distribution of molecular gas masses in low luminosity radio galaxies, in 
comparison with other radio source samples, 
confirm the suggestion that the CO is in
ordered rotation, determine its relation to the dust disks observed 
in these objects with HST and establish targets for future interferometric 
imaging.
\end{abstract}

\section{Background}   %%% Top level section head (remove "%" symbol)

The fuelling of relativistic jets in radio-loud active galactic nuclei is still
not fully understood.  It is clear that energy is released in the vicinity of a
supermassive black hole, but whether the mechanism is direct electromagnetic
extraction of rotational kinetic energy (e.g. Blandford \& Znajek 1977; 
Koide et al. 2002) or more closely related to the process of accretion 
(e.g. Blandford \& Payne 1982; Hujeirat et al. 2003) remains a matter of 
debate. The ratio of energy
flux in jets to that radiated by an accretion disk varies by orders of 
magnitude between different classes of AGN. Here, we are concerned 
with low-luminosity (FRI) radio galaxies (Fanaroff \& Riley 1974). 
These are known to contain very massive black holes, but show very little 
evidence for emission from accretion disks, their nuclear luminosities 
being a small fraction of that expected from accretion at the Eddington rate, 
$L/L_{\rm Edd} < 10^{-4}$ (Marchesini, Celotti, \& Ferrarese 2004). 
They form the parent population for nearby BL Lac objects and must
therefore produce highly relativistic jets on sub-pc scales. There is direct
evidence for relativistic motion on parsec scales in FR\,I jets (Giovannini et
al., 2001) and for smooth deceleration from relativistic to sub-relativistic
speeds on scales of 1 -- 10\,kpc (Laing et al., 1999).
FR\,I radio sources are located in fairly normal elliptical galaxies, 
invariably containing hot, X-ray emitting plasma 
(thought to confine the jets on large scales), but little ionized, 
line-emitting material at $\approx10^4$\,K. It has recently become clear that 
they may also contain substantial amounts of cool gas and dust 
(e.g. de Ruiter et al. 2002; Verdoes Kleijn \& de Zeeuw 2005; Lim et
al. 2000, 2003).  
Dust is observed in 53\% of the B2 sample of nearby radio galaxies 
(mostly FRI; see below) and the dust mass is correlated with radio power 
(de Ruiter et al., 2002). There is also a connection between the dust-lane 
morphology (disk/irregular) and the presence of jets, and some tendency for 
dust lanes and jets to be orthogonal (de Ruiter et al. 2002; 
Verdoes Kleijn \& de Zeeuw 2005). 
These associations argue that accretion of cool gas may indeed power the
radio jets.
The next step is to understand the dynamics of the cool gas. Lim et al. (2000)
detected $^{12}$CO (1 $\rightarrow$ 0) and (2 $\rightarrow$ 1) emission from 
the FR\,I radio galaxies 3C\,31 and 264 with the IRAM 30m Telescope and 
established that the line profiles indicate disk rotation. 
Interferometric observations of 3C\,31 by Okuda et al. (2005) showed
that the CO coincides spatially with the dust disk observed by HST (Martel et
al., 1999) and is in ordered rotation.  These authors suggest that the cool gas
is in stable orbits. \\
In order to increase the number of FR\,I radio galaxies with CO 
observations and so to improve our knowledge of their molecular gas properties,
we are studying a volume-limited sample of 23 nearby 
($z<0.03$) low luminosity radio galaxies, selected from the B2 
catalogue (Colla et al., 1975). We notice that for 16 of such objects HST 
imaging is available (Capetti et al., 2000).
The CO properties of this sample are compared to the 23 $z<0.03$ 
3C radio galaxies studied by 
Lim et al. (2003) and to the 18 $z<0.0233$ (or $v<7000$ km/s) 
UGC galaxies with radio jets, studied by Leon et al. (2003). 
We notice that the three samples are partially overlapping.

\begin{table}[t]
\caption{CO line measurements}
\footnotesize
\vspace{0.3truecm}
\centering
\begin{tabular}{crcrrcrr}
\hline
\multicolumn{1}{c}{Source} & \multicolumn{3}{c}{$^{12}$CO(1$\rightarrow$0)} & 
\multicolumn{3}{c}{$^{12}$CO(2$\rightarrow$1)} 
& \multicolumn{1}{c}{$\log{M_{H_2}}$}\\
\multicolumn{1}{c}{} & \multicolumn{1}{c}{$\Sigma T_a^* dv$}  & 
\multicolumn{1}{c}{$\Delta v_{\tt FWHM}$} & 
\multicolumn{1}{c}{$\frac{T_{\rm peak}}{T_{\rm rms}}$} & 
\multicolumn{1}{c}{$\Sigma T_a^* dv$}  & 
\multicolumn{1}{c}{$\Delta v_{\tt FWHM}$} & 
\multicolumn{1}{c}{$\frac{T_{\rm peak}}{T_{\rm rms}}$} & 
\multicolumn{1}{c}{} \\
\multicolumn{1}{c}{} & \multicolumn{1}{c}{K km/s}  & 
\multicolumn{1}{c}{km/s} & 
\multicolumn{1}{c}{} & 
\multicolumn{1}{c}{K km/s}  & 
\multicolumn{1}{c}{km/s} & 
\multicolumn{1}{c}{} & 
\multicolumn{1}{c}{M$\odot$} \\
\hline
$0120+33$ & $<0.35$ & - & - & $<0.61$ & - & - & $<7.7$ \\
$0149+35$ & $1.34$ & 545 & 5.4 & $3.15$ & 551 & 7.7 & $8.3$ \\
$0258+35$ & $0.88$ & 382 & 6.6 & $0.21$ & 127 & 2.4 & $8.1$ \\
$0326+39$ & $<0.35$ & - & - & $0.64$ & 171 & 1.9 & $<8.0$ \\
$0331+39$ & $1.35$ & 807 & 3.6 & $0.90$ & 679 & 3.4 & $8.5$ \\
$1122+39$ & $6.84$ & 670 & 23.0 & $5.93$ & 672 & 16.7 & $8.2$ \\
$1217+29$ & $0.58$ & 346 & 3.3 & $<0.87$ & 335 & 3.1 & $6.1$ \\
$1321+31$ & $<0.35$ & - & - & $1.24$ & 297 & 2.7 & $<7.7$ \\
$1615+35$ & $<0.35$ & - & - & $0.52$ & 129 & 4.1 & $<8.2$ \\
\hline
\multicolumn{8}{l}{$M_{H_2}$ derived from CO(1$\rightarrow$0) 
measures and assuming $H_0=100$ km/s/Mpc}
\end{tabular}
\end{table}

\section{Line Observations and Measurements} 

We used the IRAM 30m telescope to search for emission in the 
(1$\rightarrow$0) and (2$\rightarrow$1) transitions of $^{12}$CO in 9
B2 radio galaxies of the $z<0.03$ volume-limited sample, described above. 
We used receivers A100 and B100 connected to the 1 MHz filter bank in $2\times
512$ MHz blocks together with receivers A230 and B230 with the 4 MHz filter 
bank in $2\times 1$ GHz blocks. 
After averaging the outputs of the two pairs of receivers
we got noise levels of $T_{\rm rms}\sim 0.5-1$ 
mK and $\sim 1-2$ mK 
respectively ($1\sigma$, $\Delta v \sim 40$ km/s).
The data were reduced with the CLASS package and line fluxes were
measured by numerically integrating over the channels in the line profile.
Line widths were measured as full widths at 50\% of the peak. 
A source was considered detected when both (1$\rightarrow$0) and 
(2$\rightarrow$1) emission lines have $T_{\rm peak}>2 T_{\rm rms}$, 
with at least one having $T_{\rm peak}>3T_{\rm rms}$. 
In case of non detections, upper limits were 
calculated (see Evans et al., 2005). A summary of our CO line measurements 
is given in Table 1.
Line widths of the 5 detected sources are of the order of 500 km/s, 
and in a few 
cases lines show a double-horn structure, indicating rotating CO disks.
$H_2$ molecular masses were derived as in Lim et al. (2000).

\section{Molecular gas properties of the B2 $z<0.03$ sample} 

\begin{figure}[t]
\resizebox{6cm}{!}{\includegraphics{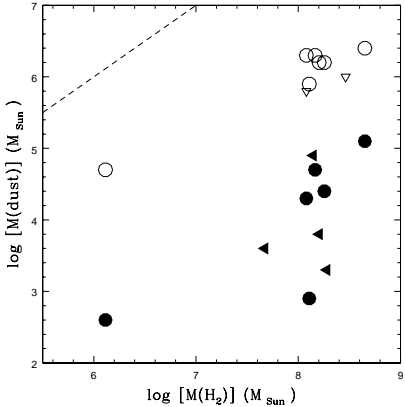}}
\hfill
\parbox[b]{70mm}{
\caption{Dust mass vs. $H_2$ mass for the $z<0.03$ B2 radio galaxies
with CO line observations. Filled symbols: $M(\rm dust)$ derived from
HST. Empty symbols: $M(\rm dust)$ derived from IRAS. 
Triangles indicate upper limits in dust mass (downwards) or
$H_2$ mass (leftwards). The dashed line indicate the 
$M(\rm dust)=M(H_2)$ locus.}}
\end{figure}

In Fig. 1 we compare the molecular mass content, $M_{H_2}$, 
to the content in dust, $M(\rm dust)$, as derived from both IRAS 
far-infrared (empty symbols) and HST optical (filled symbols) 
observations for the B2 $z<0.03$ radio galaxies. 
To our measurements (see Table 1) we have added
CO line measurements for other 7 galaxies which were observed as part of
the UGC and 3C samples (Leon et al. 2003; Lim et al. 2003, see Sect. 1). 
This means that our analysis is based on 16 of the 23 galaxies 
(i.e. $\sim 70\%$ of the whole sample). 
We notice that  
$M_{H_2}>M_{\rm dust}(IRAS)>M_{\rm dust}(HST)$. The fact that 
$M_{\rm dust}(IRAS) > M_{\rm dust}(HST)$ probably just reflects the
different scales probed in the two cases: 
the whole galaxy for IRAS FIR observations and the
inner galaxy core for HST high resolution observations. On the other
hand, the CO observations are approximately sensitive to the same scale 
as HST ($<5-10$ kpc at the $^{12}$CO(1$\rightarrow$0) observing frequency), and
a comparison is more meaningful. \\
Our observations show several hints for a physical link between the
dust component probed by HST and the molecular gas 
probed by CO: CO was detected only in those galaxies showing dust
in HST images and double-horn CO lines were found in two galaxies, both 
showing HST rotating dusty disks (see example in Fig. 2). 
Such evidences reinforce previous indications by Lim et al. (2000) and
Okuda et al. (2005). For further confirmation on the gas dynamics, the most 
suitable B2 radio sources will be proposed for interferometry at Plateau de 
Bure.

\begin{figure}[t]
\centering
\begin{minipage}[c]{0.5\textwidth}
\centering
\includegraphics[width=5.5cm,angle=-90]{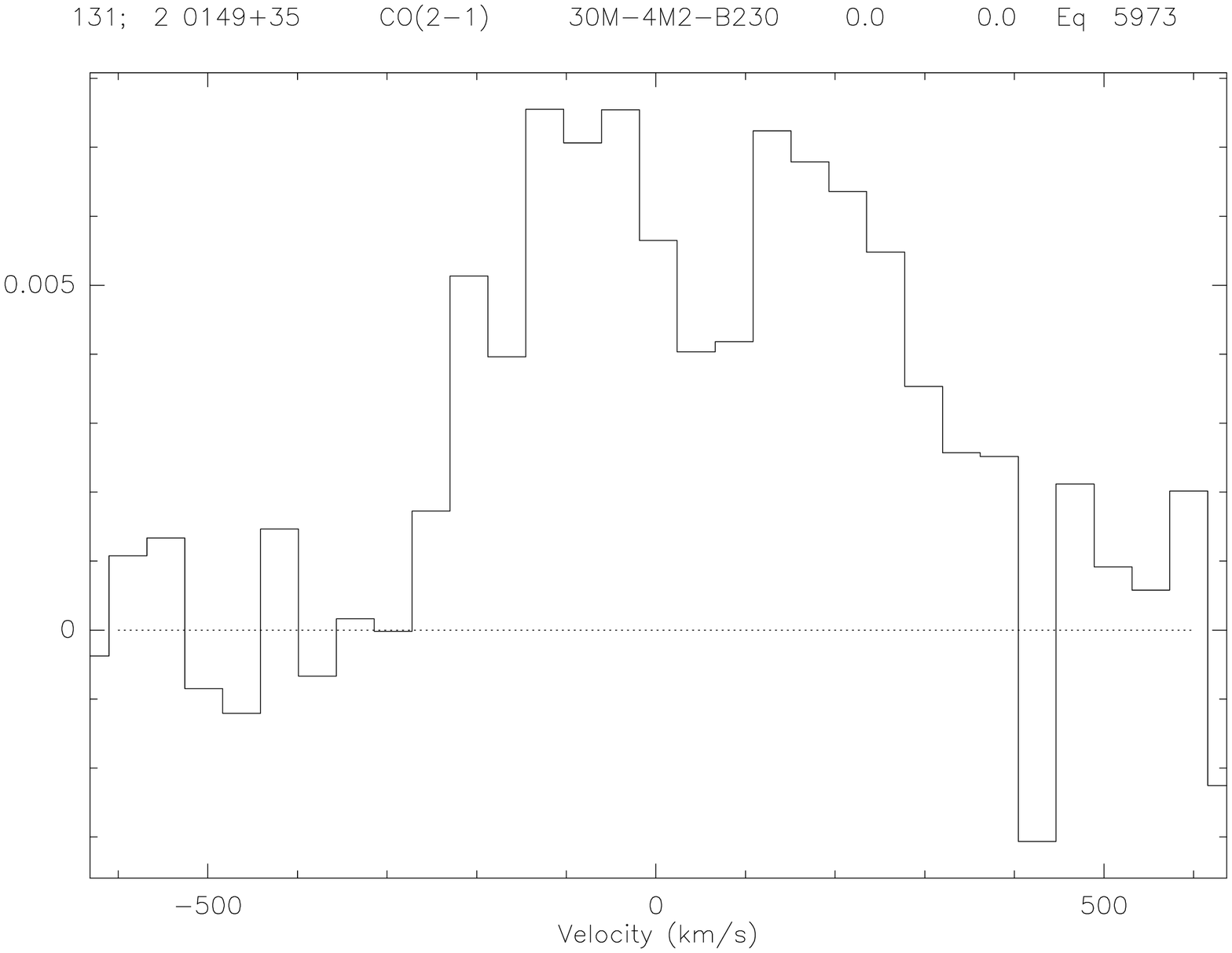}
\end{minipage}%
\begin{minipage}[c]{0.5\textwidth}
\centering
\includegraphics[width=5.5cm]{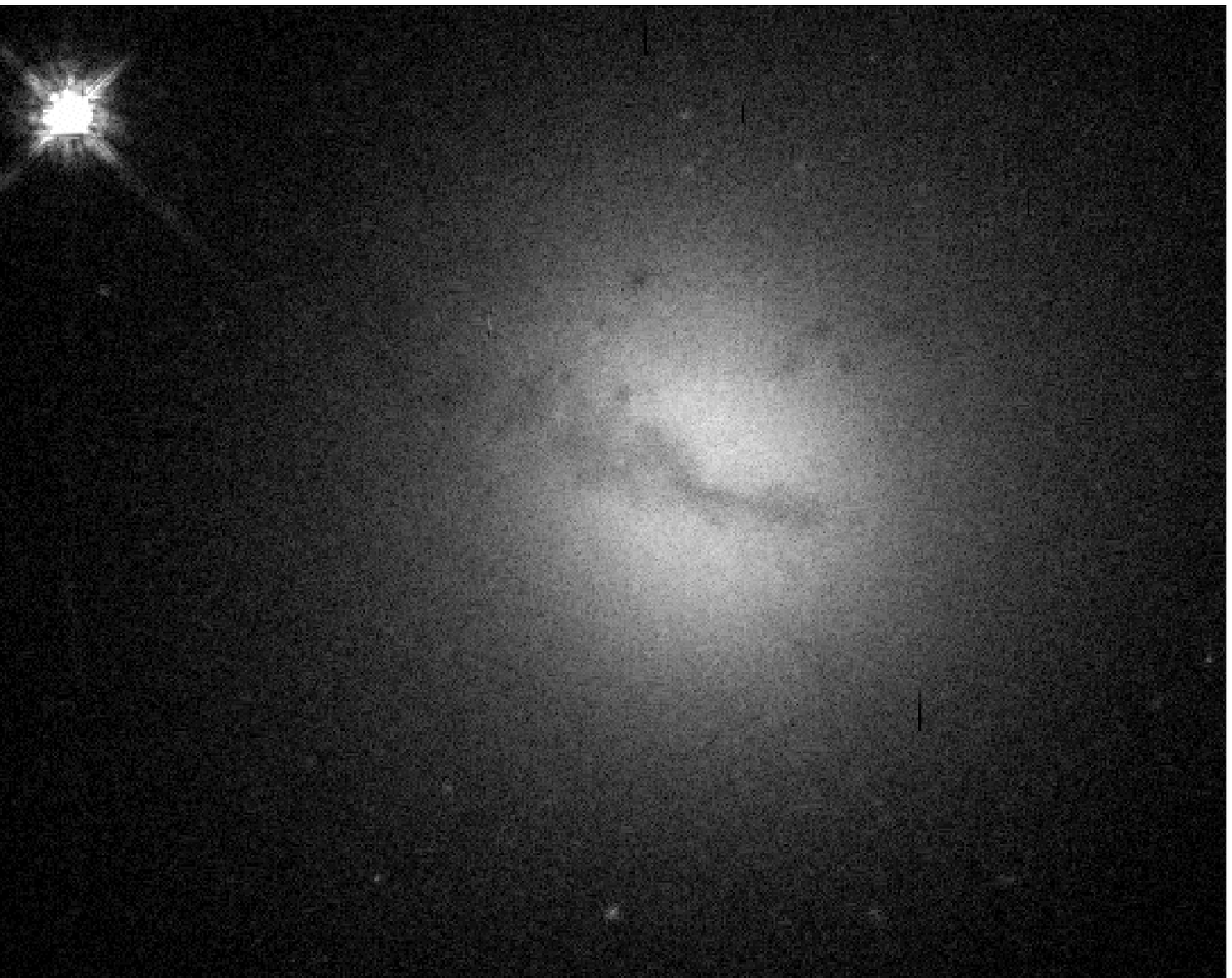}
\end{minipage}
\caption[]{{\it Left:} $^{12}$CO(2$\rightarrow$1) spectrum obtained at IRAM 
30m for the B2 radio galaxy 0149+35, showing a double-horn line. {\it Right:} 
HST image of 0149+35, showing a dusty rotating disk.}
\end{figure}

\section{Comparison with other samples} 

We compared the B2 $z<0.03$ radio galaxy sample to the $z<0.03$  3C and the 
$z<0.023$ UGC radio samples (Lim et al. 2003; Leon et al. 2003).
To take into account the differences in volume, 
we limited the comparison to all objects with $z \leq 0.023$. In such range
we found no significant difference: molecular 
gas masses are very similar, approximately spanning the 
range $10^7 - 10^9$ M$_\odot$, with upper limits varying 
from $\sim 10^7$ M$_\odot$
to $\sim 10^8$ M$_\odot$, depending on distance. Also CO detection rates 
do not differ significantly in the three samples: $47\pm 17\%$, 
$33\pm 17 \%$, and $75\pm 25 \%$,  for the UGC, 3C and B2 samples respectively.
We notice however that the statistics is poor and larger samples are needed
to better constrain the molecular gas properties of such objects.

%\subsection{}   %%% Second level section head (remove "%" symbol)
%\subsubsection{}   %%% Lowest level section head (remove "%" symbol)
%\section*{}    %%% Unnumbered top level section head (remove "%" symbol)
%\subsection*{}   %%% Unnumbered second level section head (remove "%" symbol)

%\acknowledgements %%% Text of acknowledgements runs on after this command.

%%% THE BIBLIOGRAPHY
%%%
%%% CONSULT SECTION 3 OF "INSTRUCTIONS FOR AUTHORS" FOR HOW TO USE NATBIB.
%%% AUTHORS ARE ENCOURAGED TO USE EITHER THE "THEBIBLIOGRAPY" ENVIRONMENT
%%% BY UNCOMMENTING (DELETING THE "%" SYMBOL) THE COMMANDS BELOW, OR BY
%%% USING THE BIBTEX ENVIRONMENT. TO FIND OUT WHICH IS APPLICABLE TO YOUR
%%% CONTRIBUTION, CONSULT THE VOLUME EDITORS FOR YOUR PROCEEDINGS.
%%%

\end{document}